\documentclass[sigconf]{acmart}
\usepackage{array}

\AtBeginDocument{%
  \providecommand\BibTeX{{%
    \normalfont B\kern-0.5em{\scshape i\kern-0.25em b}\kern-0.8em\TeX}}}

\setcopyright{acmcopyright}
\copyrightyear{2018}
\acmYear{2018}
\acmDOI{XXXXXXX.XXXXXXX}

\acmConference[The Web Conference 2024]{Singapore}
\acmPrice{Singapore}
\acmISBN{978-1-4503-XXXX-X/18/06}

\begin{document}

\title{Semantic Ads Retrieval at Walmart eCommerce with Language Models Progressively Trained on Multiple Knowledge Domains}

\author{Zhaodong Wang}
\authornote{Both authors contributed equally to this research.}
\authornote{Corresponding authors.}
\email{zhaodong.wang@walmart.com}
\author{Weizhi Du}
\orcid{0000-0001-7448-8190}
\authornotemark[1]
\authornotemark[2]
\email{weizhi.du@walmart.com}
\author{Md Omar Faruk Rokon}
\orcid{0000-0002-1385-9389}
\authornotemark[1]
\email{mdomarfaruk.rokon@walmart.com}
\affiliation{%
  \institution{Walmart Global Tech}
    \streetaddress{ 860 W California Ave.}
  \city{Sunnyvale}
  \state{California}
  \country{USA}
  \postcode{94086}
}

\author{Pooshpendu Adhikary}
\author{Yanbing Xue}
\author{Jiaxuan Xu}
\email{pooshpendu.adhikary@walmart.com}
\email{yanbing.xue@walmart.com}
\email{Jiaxuan.Xu@walmart.com}
\affiliation{%
  \institution{Walmart Global Tech}
    \streetaddress{ 860 W California Ave.}
  \city{Sunnyvale}
  \state{California}
  \country{USA}
  \postcode{94086}
}
\author{Jianghong Zhou}
\author{Kuang-chih Lee}
\author{Musen Wen}
\email{jianghong.zhou@walmart.com}
\email{Kuangchih.Lee@walmart.com}
\email{musen.wen@walmart.com}

\affiliation{%
  \institution{Walmart Global Tech}
    \streetaddress{ 860 W California Ave.}
  \city{Sunnyvale}
  \state{California}
  \country{USA}
  \postcode{94086}
}


\renewcommand{\shortauthors}{Zhaodong Wang and Weizhi Du, et al.}
\begin{abstract}
Sponsored search in e-commerce poses several unique and complex challenges. These challenges stem from factors such as the asymmetric language structure between search queries and product names, the inherent ambiguity in user search intent, and the vast volume of sparse and imbalanced search corpus data. The role of the retrieval component within a sponsored search system is pivotal, serving as the initial step that directly affects the subsequent ranking and bidding systems. In this paper, we present an end-to-end solution tailored to optimize the ads retrieval system on Walmart.com. Our approach is to pretrain the BERT-like classification model with product category information, enhancing the model's understanding of Walmart product semantics. Second, we design a two-tower Siamese Network structure for embedding structures to augment training efficiency. Third, we introduce a Human-in-the-loop Progressive Fusion Training method to ensure robust model performance. Our results demonstrate the effectiveness of this pipeline. It enhances the search relevance metric by up to 16\% compared to a baseline DSSM-based model. Moreover, our large-scale online A/B testing demonstrates that our approach surpasses the ad revenue of the existing production model. 
\end{abstract}

\ccsdesc[500]{Information systems~Information retrieval}
\ccsdesc[300]{Applied computing~E-commerce}
\ccsdesc{>Computing methodologies~Machine learning approaches}
\ccsdesc[100]{Information systems~Clustering and classification}

  
\keywords{Retrieval, Search, BERT, E-commerce Search, language models}


\maketitle

\section{Introduction}
In the digital age, the landscape of e-commerce has rapidly evolved, with search advertising emerging as a vital component of online marketplaces. Platforms like Walmart.com, which cater to millions of users daily, rely increasingly on sophisticated advertising mechanisms to drive their revenue streams \cite{xue2023practical}. The strategic display of advertisements, or sponsored products, is not merely a supplement to the user experience but a critical driver of customer engagement and sales. The motivation behind enhancing ad relevance is twofold: first, to improve the user experience by ensuring that customers are presented with ads that are closely aligned with their search intent, thereby fostering a more intuitive and satisfying shopping journey. Second, there is a direct economic incentive, as more relevant ads are likely to lead to higher click-through rates, culminating in increased ad-generated revenue. This is particularly significant for e-commerce giants like Walmart, where small percentage points in ad performance can translate to substantial financial outcomes.

The central challenge in search advertising is not just the real-time delivery of ads, but more importantly, the relevancy of these ads to the users' search queries. In an online marketplace as vast as Walmart.com, which serves millions of users every day, presenting sponsored products that precisely match user intent is critical. Relevant ads enhance the shopping experience by providing users with options that align with their needs, thereby increasing the click-through rates and likelihood of purchases. This relevancy is not only beneficial for customer satisfaction, but also a key driver of the economic success of e-Commerce platforms, where even minor improvements in ad relevance can lead to significant increases in ad-generated revenue.

Historically, the retrieval module of traditional search ad systems has faced significant hurdles. It must sift through all available ad products to find potential candidates, which are then ranked based on predicted Click-Through-Rate (pCTR) and bidding prices. The effectiveness of this retrieval module is foundational to the overall ad system's performance. Yet, despite its critical role, there is a notable lack of comprehensive guides available to industry professionals on creating and implementing end-to-end retrieval pipelines that can address the associated challenges effectively.

In the early stages of item retrieval, roughly two decades ago, collaborative filtering (CF) technology\cite{schafer2007collaborative} held sway but grappled with the challenges posed by data sparsity, cold starts, and the rapidly expanding scale of data in e-Commerce. Other classic retrieval techniques, such as content-based (CB) and BM-25 (TF-IDF and token matching)\cite{BM25}, required a lot of domain knowledge to handcraft features and struggled to generalize to new items or downstream tasks. However, in the past decade, significant progress in deep learning has revolutionized retrieval models through embedding-based approaches, demonstrating their robust representation-learning capabilities\cite{alibaba}. In 2013, Huang et al. introduced Deep Structured Semantic Models (DSMM)\cite{dssm} for the retrieval of web searches, establishing the foundation for subsequent innovations. In 2016, Covington et al. pioneered the Two-Tower embedding architecture for recommending YouTube videos\cite{youtube}. One year later in 2017,  Facebook's introduction of Faiss\cite{faiss} offered an efficient framework for retrieving document embeddings that are similar to each other by searching for the approximate nearest neighbor (ANN). Since then, the two-tower architecture that integrates dual embedding networks and nearest neighbor search has become widely accepted in different retrieval and recommendation scenarios\cite{ebay}\cite{youtube2}. In recent years, there has been significant progress and success in the application of deep learning models to various domains, such as speech recognition \cite{kamath2019deep}, computer vision \cite{du2022self}\cite{huang2022unsupervised}, and others \cite{du2021improved}. As a result, the adoption of semantic embedding based on deep learning models has emerged as the mainstream choice in many industries. The revelation of Bert (Bidirectional Encoder Representations from Transformers)\cite{2018bert} in 2018 sent shock waves through the natural language processing community, breaking records, and setting new standards. Numerous studies have demonstrated that substituting the original embedding networks with Bert, the Twin-Tower\cite{TwinBERT}(or its derived variant \cite{Quadruplet}) architecture, can yield exceptional results in information retrieval tasks. In 2022, A. Magnani, et al. proposed a novel model designed to enhance the organic search retrieval system\cite{magnani2022semantic}. However, it was observed that this model did not significantly contribute to the improvement of the advertising system on Walmart.com.

In response to the evolving landscape of e-commerce search advertising, we introduce a holistic end-to-end pipeline that leverages the latest advancements in natural language processing and deep learning. Our novel approach, anchored by a BERT-based model, is systematically designed to unravel the complexities of e-commerce search queries, enhancing ad relevance and user experience on Walmart.com. The pipeline features:

\begin{itemize}
  \item A pre-training phase where a BERT-based classification model is refined using product categorical labels from Walmart's extensive catalog, enabling it to recognize a wide array of product categories.
  \item A two-tower Siamese network structure optimized for cosine similarity, ensuring that retrieved ads closely align with the semantic content of user queries.
  \item Progressive training across diverse knowledge domains, which equips the model with a nuanced understanding of natural language and Walmart products.
  \item The deployment of a high-throughput online inference and retrieval system capable of operating at the scale demanded by Walmart's substantial e-Commerce traffic.
\end{itemize}

This comprehensive approach not only advances the technical framework for ad retrieval systems but also promises significant improvements in user engagement and satisfaction for Walmart's online platform.

The key contributions of this work are as follows:


\begin{itemize}
\item We introduce a novel two-stage progressive training framework for our retrieval embedding model, enhancing its capability for nuanced e-Commerce comprehension.

\item We develop a unique fusion knowledge training approach with human-in-the-loop mechanism for dynamic weight adjustment, utilizing a Siamese network architecture, to attain both a broad semantic understanding and targeted e-Commerce expertise.

\item We present a new method for hard negative data labeling, demonstrating its effectiveness in improving retrieval model performance.

\item We show that our approach improves both the relevance and business metrics with a substantial margin compared to the baseline approach.

\item We illustrate the scalability and robustness of our solution, highlighting its successful deployment in a real-world environment, supporting one of the world's largest e-Commerce platforms.
\end{itemize}

\section{Model Architecture AND TRAINING METHODOLOGY}
In this section, we explore the structural design and training methodologies of the language models in use. Our discussion highlights the utilization of BERT models for their superior text embedding capabilities and outlines our approach to training these models with a curated mix of datasets tailored to the e-Commerce domain. Furthermore, we detail the architecture of the Siamese network and describe the training process, which includes a human-in-the-loop mechanism for dynamic weight adjustment based on model performance feedback.

\begin{figure*}
  \centering
  \includegraphics[width=\linewidth]{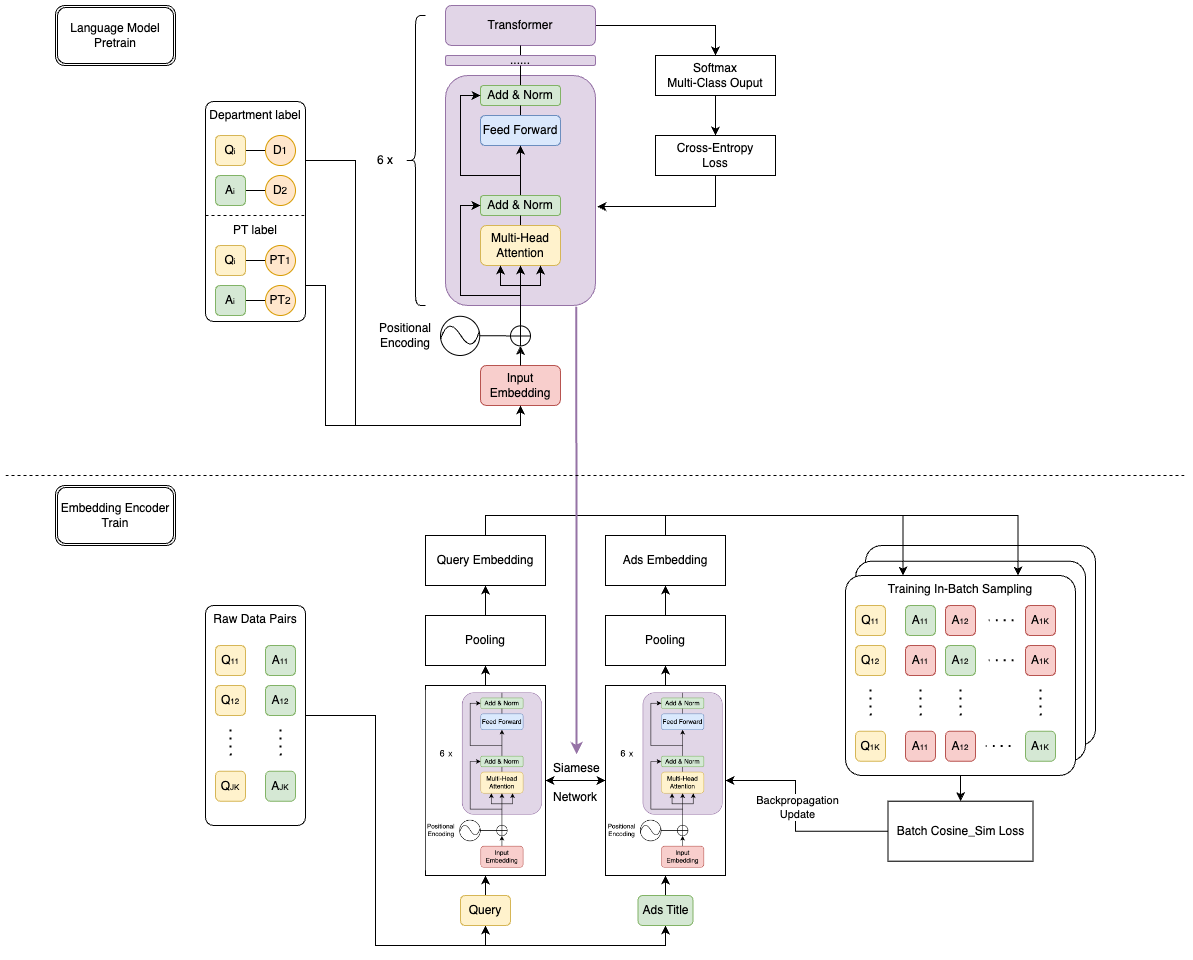}
  \caption{Two-stage progressive training architecture of retrieval embedding model.}
  \Description{Two-stage progressive training architecture of retrieval embedding model.}
  \label{fig:model_architecture}
\end{figure*}

\subsection{Models}
We utilize BERT's language model, acknowledged for producing meaningful text embeddings \cite{2018bert}. To suit our efficiency needs, DistilBERT \cite{2019distilbert}, a compact version of BERT, is employed for its balance of performance and speed. Nonetheless, the generic BERT model's training on broad corpora is insufficient for niche applications like e-Commerce, necessitating a tailored approach for sentence pair regression tasks due to their combinatorial complexity.

\begin{figure*}
  \centering
  \includegraphics[width=0.9\linewidth]{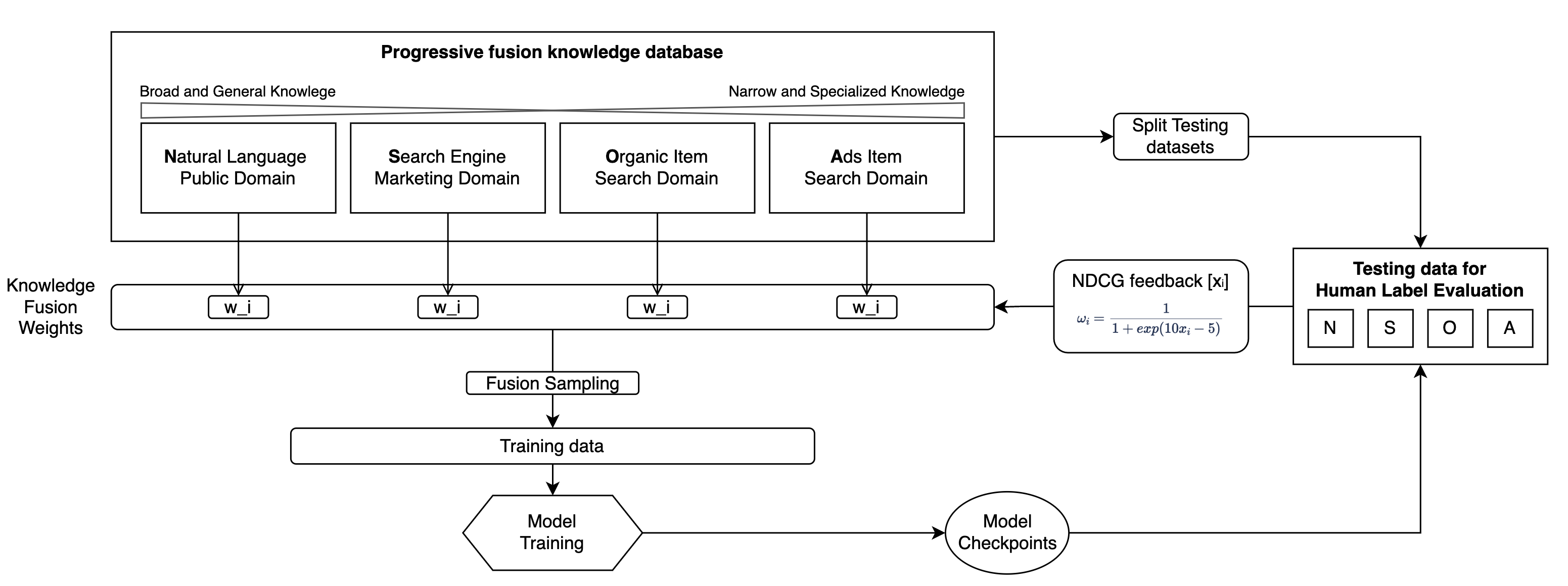}
  \caption{Human-in-the-loop knowledge fusion sampling.}
  \Description{Human-in-the-loop knowledge fusion sampling.}
  \label{fig:model_sample}
\end{figure*}

\textbf{\textit{Model Training Architecture:}}
We propose a two-stage progressive training architecture for our retrieval embedding model consisting of 1) language model pre-training, and 2) embedding encoder training, as illustrated in Figure~\ref{fig:model_architecture}. In the initial stage of language model pretrain, DistilBERT is utilized to execute multiple multi-class classification tasks. These tasks leverage Walmart's product category data including 'Department' labels and 'Product Type' labels, assigning specific categories to each item or query on Walmart.com. Given the fixed and unvarying nature of the 'Product Type' and 'Department', a multi-class classification approach is deemed appropriate. Detailed dataset information is available in Section 2.2. Cross-entropy loss serves as the pretrain objective function, effectively leveraging Walmart's categorical label insights into the BERT transformers. In the subsequent phase, we incorporate a pooling layer atop the pre-trained DistilBERT output, generating a 384-dimensional sentence embedding. This embedding model is then integrated into a Siamese network to directly train sentence pairs on optimizing cosine similarity loss.

During the second stage of training, we incorporate a variety of knowledge domains through a process of progressive fusion training. This strategy ensures that the model not only has the ability of general semantic understanding but also acquires a deep understanding of the e-Commerce knowledge. Diverse datasets facilitate the learning across these multiple domains as shown in Table 1. We designed and implemented a framework with the human-in-the-loop dynamic fusion sampling mechanism, as shown in Figure~\ref{fig:model_sample}. This system allows for the recalibration of training and sampling weights based on the assessment of human-labeled data. Weights adjustments are governed by the equation:

\begin{equation}
w_i = \frac{1}{1+exp(10*x_i-5)},
\end{equation}

where $w_i$ represents the sampling weights for each domain, scaled between 0 and 1, and $x_i$ denotes the normalized discounted cumulative gain (nDCG) scores calculated from human-labeled data. It is evident that as the nDCG scores increase, indicating proficient performance, the corresponding domain weights decrease. This inverse relationship allows our model to de-prioritize domains where performance is already satisfactory and focus more on other domains requiring improvement.

For the encoder training, we adopt a direct approach using cosine-similarity labels, which enables us to formulate the loss function based on the mean squared error (MSE) loss as follows:
\begin{equation}
L(V_x,V_y,l) = (V_x*V_y - l)^2,
\end{equation}
where $V_x$ and $V_y$ are the input sentence pairs and $l$ is ground-truth label. To enhance the model's discriminative learning, we apply in-batch negative sampling, refining the MSE loss to a more sophisticated triplet loss for the Walmart dataset. This approach concurrently processes positives and negatives against the anchor:

\begin{equation}
L(A,P,N) = max(D(A-P)-D(A-N)+\alpha,0),
\end{equation}

where $A, P$, and $N$ denote embeddings for anchor (query), positive (relevant item), and negative (irrelevant item) sentences. $D(A, P)$ is the L2 distance between the anchor and the positive sentence embeddings, $D(A, N)$ is the L2 distance between the anchor and the negative sentence embeddings, and the margin $\alpha$ ensures that $P$ is at least $\alpha$ closer to $A$ than $N$.

\textbf{\textit{Benchmark Models:}}
Our model is benchmarked against existing models - Deep Structured Semantic Model (DSSM) \cite{dssm}. The performance metrics are meticulously compared, underscoring the superior capabilities of our trained BERT-based model in capturing semantic textual similarities. These comparisons and findings are described in Section 4.

\begin{figure}[H]
  \centering
  \includegraphics[width=\linewidth]{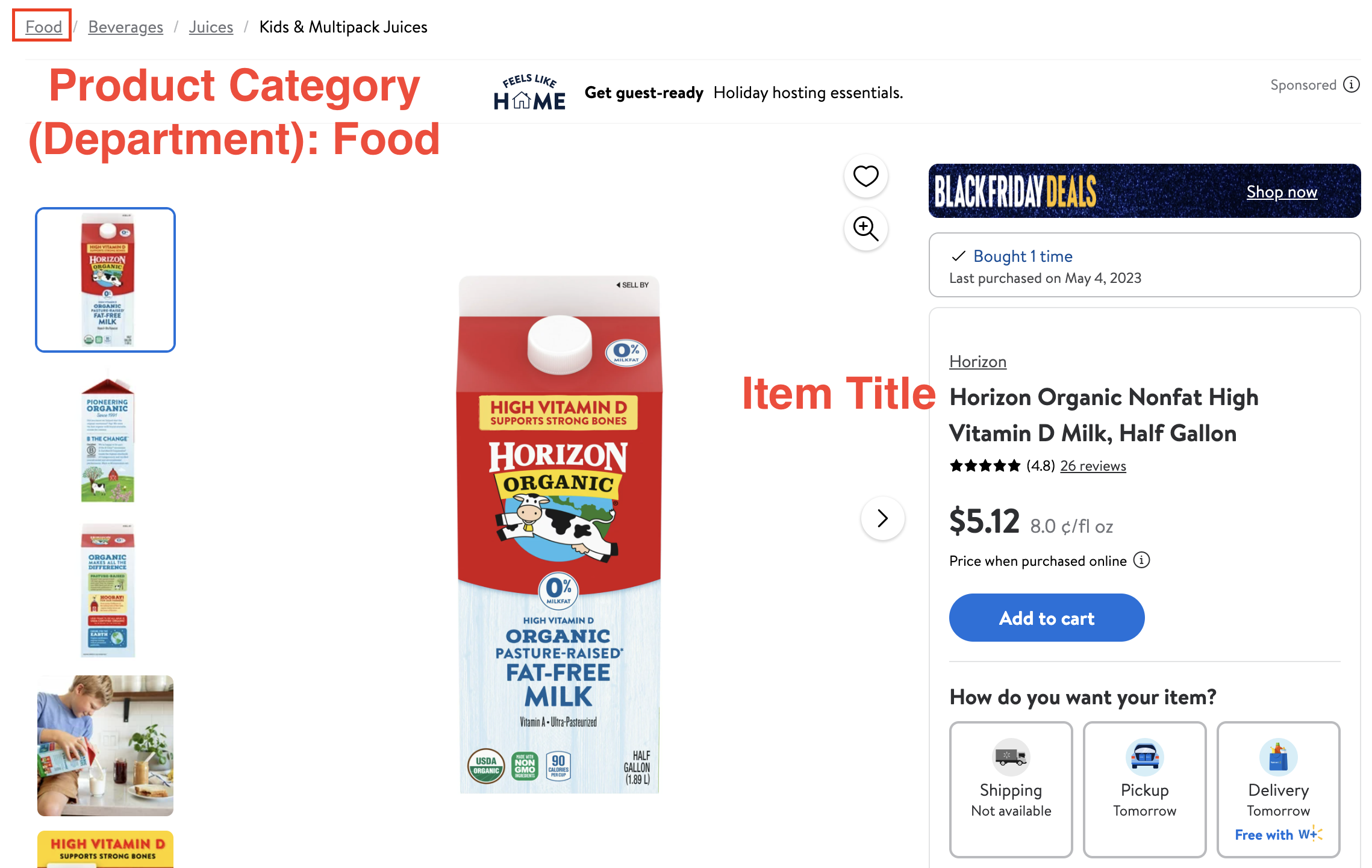}
  \caption{Department classification label of exampled item}
  \Description{.....}
  \label{fig:super_dep}
\end{figure}
\subsection{Datasets}


Our training methodology also benefits from a mix of pre-labeled open-source datasets and Walmart's proprietary log data. We establish a solid database using 200,000 sampled query - 'Department' and item - 'Department' pairs, as exemplified by the data from Walmart.com depicted in Figure~\ref{fig:super_dep}. For instance, the 'Department' for the item 'Horizon Organic Nonfat High Vitamin D Milk, Half Gallon' is classified under 'Food'. Employing the same methodology, we construct another 200,000 query - 'Product Type' and item - 'Product Type' combinations derived from Walmart's extensive internal datasets. Each query or item is uniquely paired with a corresponding product type.

For subsequent training, we collected five distinct datasets to cover four knowledge domains: 

1. the Stanford Natural Language Inference (SNLI) dataset \cite{bowman2015large}, an open-source dataset of 570k labeled sentence pairs, provides the fundamental base, offering a broad spectrum of sentence pairs annotated with logical relationships vital for initial language inference training. 

2. the Multi-Genre Natural Language Inference (MultiNLI) corpus \cite{williams2017broad} introduces complexity with sentence pairs from various genres, pushing the model's ability to process diverse linguistic contexts.

3. a specialized Search Engine dataset, comprised of 20 million query-item title pairs, is extracted from Google's search results for queries that specifically request pages from "site:walmart.com". This ensures that the dataset is reflective of Walmart's product relevance as determined by Google's search algorithms.

4. the Human Evaluated Walmart Search Log dataset, with 8 million query-item title pairs, involves human evaluators' assessment of the relevance, providing a nuanced understanding of search-result appropriateness within the Walmart context.

5. Walmart.com Ad Log Data, encompassing six months of user interactions, offers a comprehensive view of customer engagement and product preferences on Walmart's digital platform.

\newcolumntype{C}[1]{>{\centering\arraybackslash}p{#1}}

\begin{table}[h]
\centering
\caption{Summary of Datasets Used in Training}
\label{tab:datasets}
\begin{tabular}{|C{1.5cm}|C{3cm}|C{3cm}|}
\hline
\textbf{Domain} & \textbf{Dataset Name} & \textbf{Size} \\
\hline
{Natural Language Public Domain} & Stanford Natural Language Inference (SNLI) & 570k sentence pairs \\
\cline{2-3}
 & Multi-Genre Natural Language Inference (MultiNLI) & 430k sentence pairs \\
\hline
Search Engine Marketing Domain & Walmart-Focused Google Search Data & 20M query-item title pairs \\
\hline
Organic item Search Domain & Human Evaluated Walmart Search Logs & 8M query-item title pairs \\
\hline
Ads item Search Domain & Walmart.com Ads Log Data & 12M query-item title pairs \\
\hline
\end{tabular}
\end{table}

\textbf{\textit{Data Labeling for Ad Logs:}}
For Walmart's  Ads log data, we adopt a pseudo-labeling approach due to the impracticality of manually annotating the vast number of query-item interactions. This method infers relevance from user engagement signals, specifically click-through data. An item is considered relevant if it attracts a high number of clicks within a designated period, leveraging the assumption that user interaction signifies relevance.

\textbf{\textit{Negative QIP Identification:}}
Incorporating negative Query-Item Pairs (QIPs) is essential for accurately determining relevance. Our methodology distinguishes between two categories of negatives. 'Easy' negatives are those randomly chosen from categories that have no relation to the query. 'Hard' negatives, on the other hand, are derived from search logs, pinpointing items that are often displayed yet seldom receive clicks. By training the model to recognize these varied levels of relevance, we significantly improve the precision of search and recommendation systems. 

\section{Semantic Retrieval Online Service Architecture}
\label{sec:system_vespa}
In this section, we will introduce the system architecture for online service deployment of semantic retrieval.

\begin{figure}[H]
  \centering
  \includegraphics[width=1.0\linewidth]{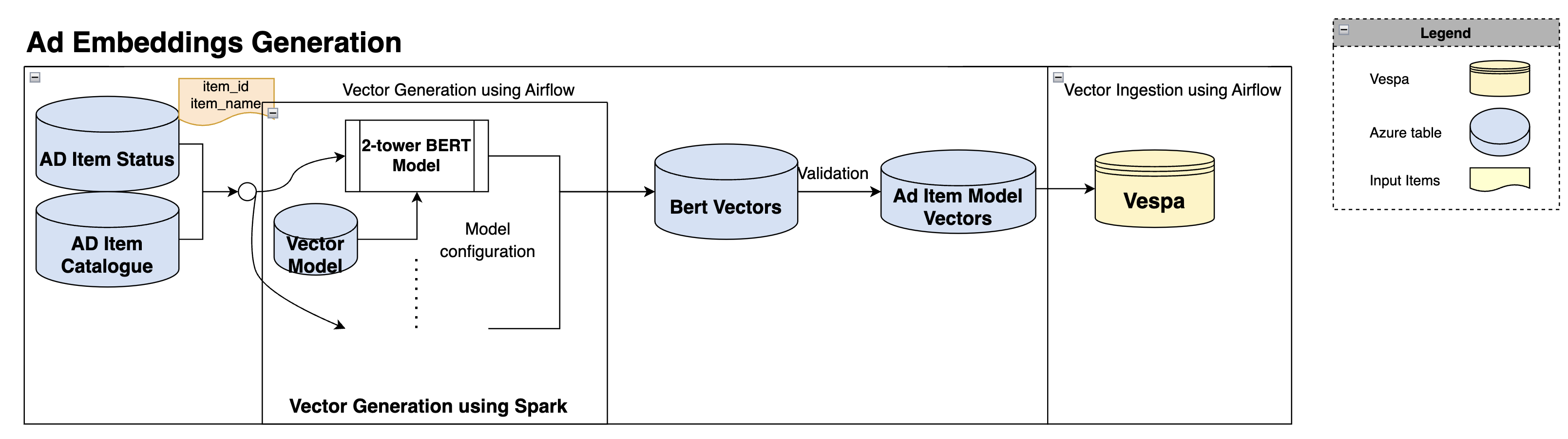}
  \caption{Pipeline architecture of generating Ads embedding}
  \Description{.....}
  \label{fig:embedding_pipeline}
\end{figure}

The fundamental data source of retrieval service is the embedding generation pipeline, which keeps generating the embedding vector of sponsored products (ads) and customers' search queries using our two-tower Bert model. Taking the ad embedding generation as an example, we show the architecture diagram in Figure~\ref{fig:embedding_pipeline}. There are two main Airflow modules: The first one keeps scanning and fetching eligible ads and distributes them onto hundreds of Spark clusters which host the two-tower Bert model, where the corresponding embedding vectors are generated thereafter. After the vector data passes the validation phase, the second Airflow module will grab the embedding vectors and ingest them into the target fields of Vespa (the vector search engine proposed by Yahoo).

\begin{figure}[H]
  \centering
  \includegraphics[width=1\linewidth]{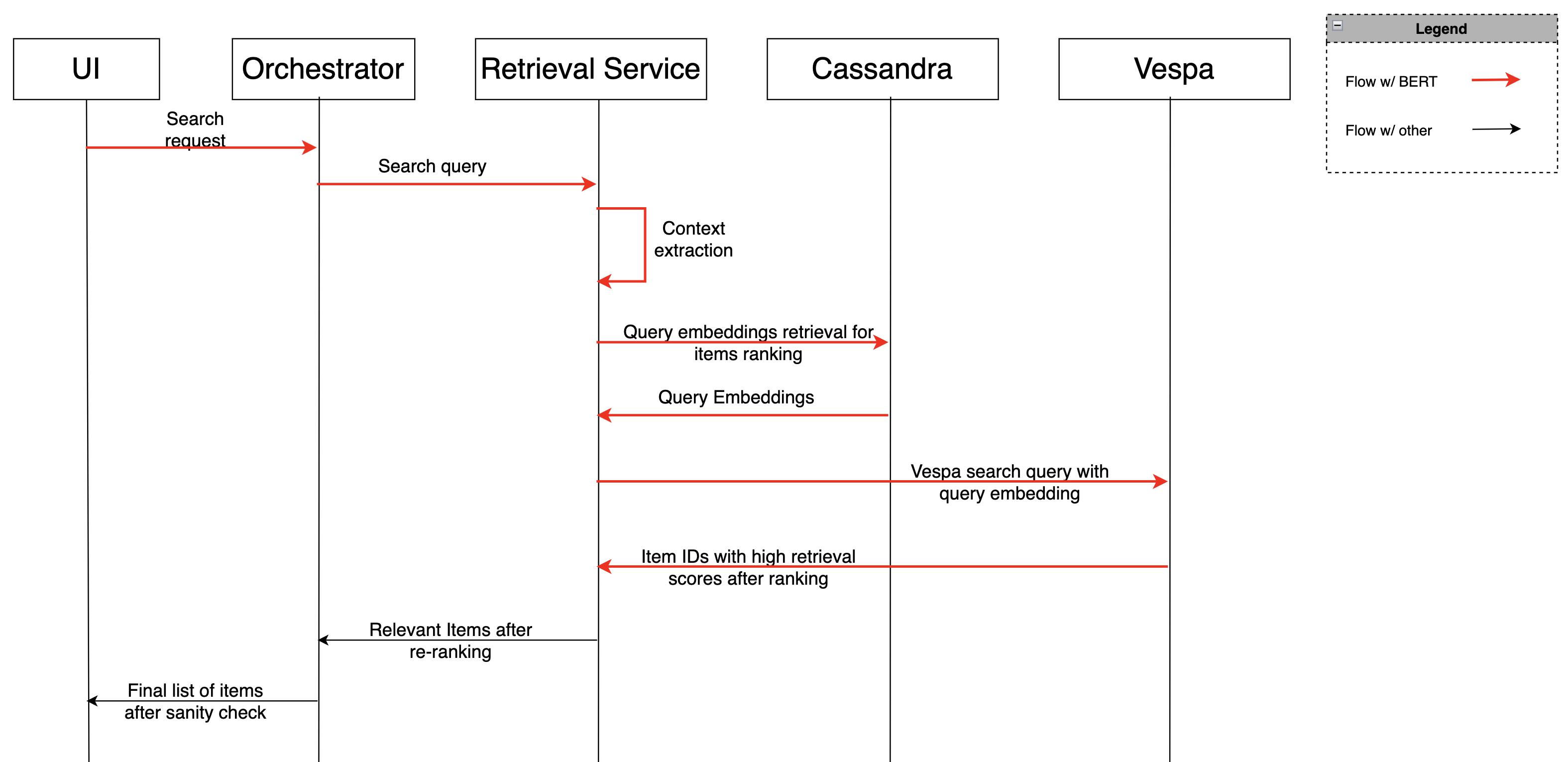}
  \caption{Retrieval service architecture based on Vespa search engine}
  \Description{.....}
  \label{fig:retrieval_service}
\end{figure}

Once we have ads and query embedding vectors streaming into the high-speed vector search engine (Vespa), we can build the real-time ads search service on top of that. The response of a search query can be divided into multiple stages as shown in Figure~\ref{fig:retrieval_service}. Given a customer's search query, Orchestrator will first process the raw query and send the processed query to Retrieval Service, where we fetch the saved query embedding vector. The embedding vector is wrapped in a designed API format to do an approximate nearest neighbor search in Vespa. Retrieved items will go through the re-rank and sanity check stages to be displayed on the customer's shopping page.

\section{Experiments and results}
In this section, we will discuss the experiment results of comparing our two-tower semantic model with the baseline DSSM model. Due to the company's confidentiality policy, we only reveal relative improvement numbers. 

\subsection{Relevance Evaluation Metrics}
The two primary matrices are used here to evaluate the relevance of the models: Normalized Discounted Cumulative Gain (NDCG) and Irrelevant Ads Rate (IAR). The metrics are defined as follows:

\textbf{\textit{Normalized Discounted Cumulative Gain (NDCG):}}  
NDCG~\cite{jarvelin2002cumulated} is a widely-adopted relevance measurement of ordered items. We use the calculation equation as follows:

\begin{equation} 
\label{eq:DCG}
NDCG_{p} = \frac{DCG_{p}}{IDCG_{p}}
\end{equation}
where DCG is defined as:
\begin{equation} 
\label{eq:DCG}
DCG_{p} = \sum_{i=1}^{p}\frac{rel_{i}}{\log_{2}(i+1)}
\end{equation}
where p is the length of item list, $rel_{i}$ is the human score of $i\_th$ item. To make DCG scores fair across different item lists, we always clip to the same list size before calculating. And IDCG is ideal (maximum) discounted cumulative gain.

\textbf{\textit{Irrelevant Ads Rate (IAR):}}  
IAR is the percentage of irrelevant ads when we look at the first n items of a retrieval list. In particular, we want to decrease this percentage when we are looking at the first few items (e.g. top 3 or top 5), otherwise, there will be a high probability that such front irrelevant items get selected by following the ranker model due to higher retrieval scores. IAR is calculated as follows:

\begin{equation} 
\label{eq:IAR}
\frac{\sum_{i=1}^{n}I(i)}{n}
\end{equation}
where for $n$ items $I(i) = 1$ if $i\_th$ item is irrelevant else it is 0.
IAR is used to directly measure a retrieval model's ability to avoid irrelevant items.

Alongside these relevance-focused metrics, additional e-Commerce indicators are also considered, such as:

\begin{itemize}
\item CTR (Click-Through Rate): The ratio of clicks to impressions on an ad.
\item CPMV (Cost Per Thousand Views): The advertising cost per thousand views.
\item Ads Revenue: The total revenue generated from ad clicks.
\end{itemize}

\subsection{Offline simulated relevance evaluation}
\begin{figure*}
  \centering
  \includegraphics[width=0.8\linewidth]{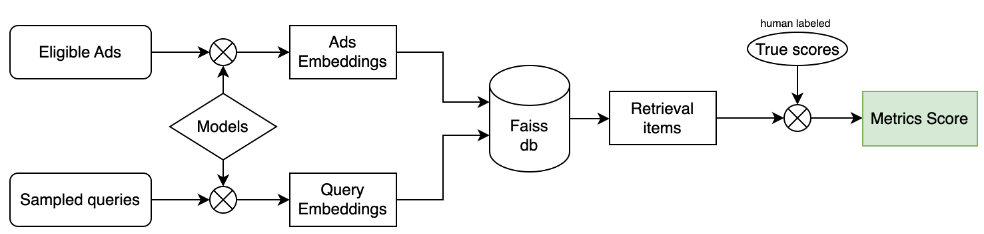}
  \caption{Diagram of Faiss-based simulation relevance evaluation}
  \label{fig:faisseva}
\end{figure*}

To perform offline retrieval simulations, we have developed offline Faiss/Vespa databases with all of Walmart's eligible sponsored products. From the search history, we randomly selected two thousand queries according to Walmart traffic segment and utilized their embeddings to retrieve 20 items per query through an approximate nearest-neighbor search in the Faiss database. The methodology of this evaluation is shown in Figure~\ref{fig:faisseva}.

For the evaluation, we collected the ground-true relevance labels of the dataset from a third-party human annotation platform specializing in annotating Walmart traffic data. They labeled query-item pairs using the three-level class method. \textit{\textbf{0}: Irrelevant; \textbf{1}: Partially Relevant; \textbf{2}: Fully Relevant}. The results of this offline evaluation are shown in Table ~\ref{tab:ndcg_iar_offline}. We observed an increase in NDCG up to 16\%, along with a decrease in IAR by about 4\%. Furthermore, our analysis revealed an average drop of 3.6\% in NDCG when not using hard negative training datasets, indicating the effectiveness of our proposed hard negative data labeling approach in enhancing model performance.

\begin{table}[H]
\caption{Offline results of Siamese network and
DSSM model}
  \label{tab:ndcg_iar_offline}
\begin{tabular}{ccc}
\toprule
Metrics Type & \begin{tabular}[c]{@{}c@{}}Siamese BERT network  \\ vs.\\  DSSM\end{tabular} \\ \hline
NDCG@5      &  \textbf{+16.10\%}                                                               \\
NDCG@10     &  +13.26\%                                                               \\
NDCG@20     & +13.42\%                                                               \\ \hline
IAR@5      &  \textbf{-4.07\%}                                                               \\
IAR@10     & -3.45\%                                                               \\
IAR@20     & -2.97\%                                                               \\

\bottomrule
\end{tabular}
\end{table}

\subsection{Online AB testing results}
After we tested the retrieval relevance superiority offline, the next steps will be production deployment (details in Section~\ref{sec:system_vespa}) and online AB testing. This section will discuss AB testing comparing our new Two-Tower Semantic model with the current production vector model (DSSM). Since the fundamental target is to improve search-ads relevance, we first perform testing on Walmart's search pages, including:(i) search in-grid, and (ii) search carousel. Results in table ~\ref{tab:search_ab_relevance}, show the relevance improvement of the final ads impression on the two above placements. Here we consider the top 4 and top 8 NDCG because the final placement of search in-grid ads on Walmart.com contains 4 slots in a row.

Notice that the NDCG improvement numbers are not exactly the same as the offline retrieval benchmark, but the superiority of our new embedding model does hold. Since Walmart.com has multiple parallel retrieval channels and the entire ads recommendation system has a funnel-like structure containing a sequence of modules like retrieval, ranking, re-ranking, etc, the final ads impressions impact during the online AB testing would be different from what we have observed from the offline evaluation which focuses on retrieval stage.

\newcolumntype{C}[1]{>{\centering\arraybackslash}p{#1}}

\begin{table}[H]
\caption{AB testing relevance improvement of Siamese Bert network on search page}
\centering
\begin{tabular}{|c|c|c|}
\hline
Search Page Module & NDCG@4   & NDCG@8   \\ \hline
Search In-grid   & +10.91\% & +9.46\%  \\ \hline
Search Carousel  & +11.87\% & +11.73\% \\ \hline
\end{tabular}
\label{tab:search_ab_relevance}
\end{table}

\newcolumntype{C}[1]{>{\centering\arraybackslash}p{#1}}

\begin{table}[H]
\caption{AB testing E-commerce metrics improvement of Siamese Bert network on search page}
\centering
\begin{tabular}{|c|c|c|}
\hline
CTR    & CPMV   & Ads Revenue \\ \hline
+7.2\% & +4.9\% & +5.16\%     \\ \hline
\end{tabular}

\label{tab:search_ab_ecommerce}
\end{table}

Besides, through the Online AB testing, we have also testified that the improved embedding retrieval model would bring in better E-commerce metrics of the search page, which is shown in Table~\ref{tab:search_ab_ecommerce}.
We have also got a significant improvement on item page modules, which is shown in Table~\ref{tab:item_ab_ecommerce}.

\begin{table}[H]
\caption{AB testing E-commerce metrics improvement of Siamese Bert network on search page}
\centering
\begin{tabular}{|c|c|c|c|}
\hline
Item Page Modules    & CTR    & CPMV   & Ads Revenue \\ \hline
Item-top Buybox      & +1.9\% & +0.8\% & +0.6\%      \\ \hline
Item-middle Carousel & +3.2\% & +6.5\% & +6.3\%      \\ \hline
Item-bottom Carousel & +2.6\% & +4.2\% & +3.5\%      \\ \hline
\end{tabular}

\label{tab:item_ab_ecommerce}
\end{table}

\subsection{Deployment of Production Launch}
Following a series of successful A/B tests that underscored the efficacy of our two-tower BERT-based semantic retrieval model, we officially launched the system into production in June 2023. This deployment marks a key advancement in Walmart.com's search advertising, introducing a more nuanced approach to understanding and matching user queries with relevant sponsored products. The rollout was the culmination of meticulous planning and optimization, ensuring that the transition into the live environment was smooth and that the model performed in alignment with our rigorous pre-launch testing standards.

Post-deployment, we have established a rigorous monitoring protocol to continuously assess the model's performance metrics, mirroring those used during the A/B testing phase. These include measures of relevance such as Normalized Discounted Cumulative Gain (NDCG) and Irrelevant Ads Rate (IAR), as well as operational metrics like eCTR, eCPMV, latency, and throughput. The results have been promising—the model's performance has proven to be stable, and the metrics have consistently followed the positive trends observed during testing. Our ongoing commitment to monitoring and fine-tuning ensures that Walmart.com's search advertising remains responsive to user needs and business goals, reinforcing our leadership in e-commerce innovation.


\section{Conclusion and Future}

In this paper, we presented an innovative end-to-end pipeline for the semantic retrieval of sponsored products at Walmart.com. Our approach harnesses the power of a BERT-based model within a two-tower Siamese network structure, significantly enhancing ad relevance by capturing the nuanced semantics of user queries. The application of this model in a production environment demonstrates its capability to improve the user experience and contribute to the economic objectives of e-commerce platforms.

However, our approach is not without its limitations. The current reliance on pre-computing query embeddings may restrict the full spectrum of semantic search capabilities, and the system's architecture must balance the trade-off between latency and performance, dictated by constraints on model size and embedding dimensions. Recognizing these challenges, our future work is directed toward incorporating more dynamic, real-time models that can accommodate larger embedding sizes without compromising response times.

As we continue to refine our semantic retrieval system, our goal is to maintain its position as a leading solution in the e-commerce domain, pushing the boundaries of search advertising to deliver unparalleled relevance and efficiency. We welcome the research community to build upon our work, fostering advancements that will shape the future landscape of online retail.

\bibliographystyle{ACM-Reference-Format}

\end{document}